\documentclass[3p]{elsarticle}
\usepackage{amsmath,amssymb,amsfonts,a4wide,graphicx,bm,times}
\usepackage{mciteplus}
\usepackage{braket}

\usepackage{hyperref}
\hypersetup{
    colorlinks,
    citecolor=green,
    filecolor=green,
    linkcolor=blue,
    urlcolor=green
}

\makeatletter \let\old@startsection=\@startsection
\renewcommand{\@startsection}[6]
{\old@startsection{#1}{#2}{#3}{#4}{#5}{#6\mathversion{bold}}}
\makeatother

\makeatletter

\@addtoreset{equation}{section}
\makeatother

\newcommand{\dket}[1]{\ket{#1}\!\rangle}
\newcommand{\dbra}[1]{\langle\!\bra{#1}}
\newcommand{\superp}[2]{\genfrac{}{}{0pt}{}{#1}{#2}}

 \def\d{\delta}
 
 \def\Re{{\rm Re ~}}
 \def\Im{{\rm Im ~}}
 \def\p{\partial}
 
 \def\a{\alpha}
 \def\b{\beta}
 
 \def\d{\delta}
 \def\e{\varepsilon}
 
 \def\th{\theta}
 
 \def\k{\kappa}
 \def\l{\lambda}

 \def\s{\sigma}
 \def\t{\tau}
 \def\th{\theta}

 \def\G{\Gamma}
 \def\D{\Delta}

 \def\o{\omega }

\def\CN{{\mathcal{N}}}
\def\CO{{\mathcal{O}}}

\def\CS{{\mathcal{S}}}
\def\CT{{\mathcal{T}}}

\def\CY{{\mathcal{Y}}}

\def\la{\left\langle}
\def\ra{\right\rangle}

\def\implies{\quad\Rightarrow\quad}

\def\tphi{\tilde{\phi}}

\def\vphi{\varphi}

\def\CS{\mathcal{S}}

\def\qf{\mathfrak{q}}
\def\Zv{\mathcal{Z}_\text{vect.}}
\def\Zbf{\mathcal{Z}_\text{bfd.}}

\def\Zinst{\mathcal{Z}_\text{inst.}}

\def\bd{{\bar\delta}}
\def\ep{\varepsilon_+}

\def\vac{\emptyset}
\def\res{\mathop{\text{Res}}}
\def\bt{{\bar\tau}}
\def\bo{{\bar\omega}}
\def\tphi{\tilde{\varphi}}
\def\bT{{\bar T}}
\def\bf{{\bar f}}
\def\mf{\mathfrak{m}}

\title{A note on the algebraic engineering of 4D $\CN=2$ super~Yang-Mills theories\tnoteref{t1}}
\tnotetext[t1]{Preprint number: KIAS-Q18022.}
\author[rvt]{J.-E. Bourgine}
\ead{bourgine@kias.re.kr}
\author[focal]{Kilar Zhang\fnref{fn1}}
\ead{kilar.zhang@gmail.com}
\address[rvt]{Korea Institute for Advanced Study (KIAS),\\ Quantum Universe Center (QUC),\\ 85 Hoegiro, Dongdaemun-gu, Seoul, South Korea}
\address[focal]{Department of Physics, National Taiwan Normal University,\\
No. 88, Sec. 4, Ting-Chou Road, Taipei 11677, Taiwan}
\fntext[fn1]{On leave from the Institute of Theoretical Physics, Chinese Academy of Sciences, Beijing 100190, Mainland China.}

\begin{document}
\begin{abstract}
Some BPS quantities of $\CN=1$ 5D quiver gauge theories, like instanton partition functions or qq-characters, can be constructed as algebraic objects of the Ding-Iohara-Miki (DIM) algebra. This construction is applied here to $\CN=2$ super Yang-Mills theories in four dimensions using a degenerate version of the DIM algebra. We build up the equivalent of horizontal and vertical representations, the first one being defined using vertex operators acting on a free boson's Fock space, while the second one is essentially equivalent to the action of Vasserot-Shiffmann's Spherical Hecke central algebra. Using intertwiners, the algebraic equivalent of the topological vertex, we construct a set of $\CT$-operators acting on the tensor product of horizontal modules, and the vacuum expectation values of which reproduce the instanton partition functions of linear quivers. Analysing the action of the degenerate DIM algebra on the $\CT$-operator in the case of a pure $U(m)$ gauge theory, we further identify the degenerate version of Kimura-Pestun's quiver W-algebra as a certain limit of q-Virasoro algebra. Remarkably, as previously noticed by Lukyanov, this particular limit reproduces the Zamolodchikov-Faddeev algebra of the sine-Gordon model.
\end{abstract}
\maketitle
\flushbottom

\section{Introduction}
Since the pioneer work of Seiberg and Witten \cite{Seiberg1994,*Seiberg1994a}, 4D $\CN=2$ super Yang-Mills (SYM) theories have manifested deep connections with integrability. Thus, in the Euclidean background $\mathbb{R}^4$, their BPS sector is described by a classical integrable system \cite{Gorsky1995,*Donagi1995}. This system becomes quantized after the (partial) omega-deformation of the background $\mathbb{R}^2_{\e_1}\times\mathbb{R}^2$, the parameter $\e_1$ playing the role of a Planck constant \cite{Nekrasov2009}. Furthermore, the partition function of the (fully) omega-deformed theories on $\mathbb{R}^2_{\e_1}\times\mathbb{R}_{\e_2}^2$ has been computed exactly by Nekrasov using a localization technique \cite{Nekrasov2003}. These quantities receive non-perturbative (instanton) corrections. They were further shown to be equal to the W-algebras' conformal blocks describing correlation functions of Liouville/Toda integrable conformal field theories under the celebrated AGT-correspondence \cite{Alday2010,*Alba2010}. Some of these integrable aspects follow from an action of Vasserot-Schiffmann's Spherical Hecke central (SHc) algebra \cite{Schiffmann2012,*Kanno2012a,*Kanno2013} (formally equivalent to the affine Yangian of $\mathfrak{gl}_1$ \cite{Prochazka2015,Tsymbaliuk2014}), for which representations of level $m$ contains the action of the $W_m$-algebra involved in the AGT correspondence. However, the complete description of algebraic structures sustaining all the observed integrable properties is still lacking.

As it turns out, the algebraic description is better understood when the gauge theory is lifted to five dimensions by an extra $S^1$. The corresponding gauge theories have $\CN=1$ supersymmetry, and describe the low-energy dynamics of $(p,q)$-branes in type IIB string theory \cite{Aharony1997,Aharony1997a}. Alternatively, they can be obtained as the topological string amplitude on a toric Calabi-Yau threefold \cite{Aganagic2005,*Iqbal2007,*Awata2005}, and the toric diagram coincides with the $(p,q)$-brane web \cite{Leung1998}. This string theory realization is the basis of an algebraic construction, first proposed in \cite{Mironov2016,*Awata2016a}, based on the Ding-Iohara-Miki (DIM) algebra \cite{Ding1997,Miki2007}, which we call here \textit{algebraic engineering}. Representations of the DIM algebra are labeled by two integer levels, and are essentially made from two basic ones called \textit{vertical} $(0,m)$ \cite{feigin2011quantum} and \textit{horizontal} $(1,n)$ \cite{Feigin2009a}. In engineering $\CN=1$ gauge theories, a representation of level $(q,p)$ is associated to each brane of charge $(p,q)$. Thus, D5 branes are associated to vertical representations $(0,1)$, and NS5-branes (possibly dressed by $n$ D5s) to horizontal ones. A refinement of this technique has been developed in which stacks of $m$ D5-branes (with different positions) correspond to vertical representations of higher levels $(0,m)$  \cite{Bourgine2017b}. The junctions between branes are described in string theory using the refined topological vertex, which has also been written in the form of an intertwiner for the DIM algebra \cite{Awata2011}. In this way, it has been possible to associate to the brane web of linear (and also D-type, see \cite{Bourgine2017c}) quivers a certain $\CT$-operator acting on the tensor product of horizontal representations associated to external (dressed) NS5-branes \cite{Bourgine2017b}. The vacuum expectation value (vev) of this operator reproduces Nekrasov's instanton partition functions, while further insertions of DIM currents define the qq-characters \cite{NPS,*Nekrasov2015,*Kim2016,Bourgine2015c,Bourgine2016}.

In this note, we extend this algebraic construction to the case of 4D $\CN=2$ theories. For this purpose, we introduce a formal current algebra seen as a degenerate limit of the DIM algebra. This algebra has a Hopf algebra structure with a Drinfeld-like coproduct, an important feature to define intertwiners. We proceed by constructing two different types of representations, which we call again vertical and horizontal, by analogy with the DIM algebra. The vertical representation is roughly equivalent to the action of SHc in which generators act on states labeled by a set of Young diagrams. On the other hand, the horizontal representation appears to be new. It is defined in terms of twisted vertex operators built upon the modes of a free boson. Then, we solve the intertwining relation coupling these two representations, thus defining a 4D equivalent of the topological vertex. From this point, the 5D method can be readily adapted to $\CN=2$ theories, and, as an example, we construct the $\CT$-operator for the pure $U(m)$ gauge theory, and re-derive both the instanton partition function and the fundamental qq-character. This $\CT$-operator commutes with the action of the degenerate DIM algebra in a certain tensored horizontal representation. In the 5D case, the corresponding action of DIM can be decomposed into q-Heisenberg and q-Virasoro actions \cite{Mironov2016}, the latter being identify with Kimura-Pestun (KP) quiver $W$-algebra for the $A_1$-quiver \cite{Kimura2015}. A similar decomposition is observed in the degenerate case, leading to define the equivalent of KP's algebra for 4D theories. In this simple case, it is identified with a certain limit of the q-Virasoro algebra \cite{Shiraishi1995} that, surprisingly, also coincides with the Zamolodchikov-Faddeev (ZF) algebra for the sine-Gordon model \cite{Lukyanov1995}.

\section{Degenerate DIM algebra and representations}
\subsection{Definition of the algebra}
Following the analysis of the degenerate limit of the DIM algebra presented in \ref{AppA}, we introduce the following algebra obeyed by four currents denoted $x^\pm(z)$, $\psi^\pm(z)$, and a central element $c$,
\begin{align}\label{algebra}
\begin{split}
&[\psi^\pm(z),\psi^\pm(w)]=0,\quad \psi^\pm(z)\psi^\mp(w)=\dfrac{g(z-w-c\e_+/2)}{g(z-w+c\e_+/2)}\psi^\mp(w)\psi^\pm(z),\\
&\psi^+(z)x^\pm(w)=g(z-w\mp c\e_+/4)^{\pm1}x^\pm(w)\psi^+(z),\quad \psi^-(z)x^\pm(w)=g(z-w\pm c\e_+/4)^{\pm1}x^\pm(w)\psi^-(z),\\
&x^\pm(z)x^\pm(w)=g(z-w)^{\pm1}x^\pm(w)x^\pm(z),\\
&[x^+(z),x^-(w)]=-\dfrac{\e_1\e_2}{\e_+}\left[\d(z-w+c\e_+/2)\psi^+(z+c\e_+/4)-\d(z-w-c\e_+/2)\psi^-(z-c\e_+/4)\right].
\end{split}
\end{align}
This algebra depends on two parameters $\e_1$ and $\e_2$, identified with the omega-background equivariant parameters. It has been convenient to introduce the notation $\e_+=-\e_3=\e_1+\e_2$, and the function 
\begin{equation}\label{def_g}
g(z)=\prod_{\a=1,2,3}\dfrac{z+\e_\a}{z-\e_\a},
\end{equation} 
plays the role of a scattering factor. It obeys the unitarity property $g(z)g(-z)=1$. In contrast with the case of DIM, the delta function $\d(z)$ here is not the multiplicative one, but it is simply defined as $2i\pi$ times the Dirac delta function.

We will not discuss in details the mathematical definition of the algebra \ref{algebra} that should correspond to a Drinfeld double of the affine Yangian of $\mathfrak{gl}_1$ \cite{Tsymbaliuk2014}.\footnote{We would like to thank Anton Nedelin, Sara Pasquetti and Yegor Zenkevich for communicating this observation to us.} Instead, we will present two types of representations: an horizontal representation of weight $u$, denoted $\rho_u^{(H)}$ and having level $\rho_u^{(H)}(c)=1$, and a set of vertical representations of rank $m$ and weight $\vec a$, denoted $\rho_{\vec a}^{(m)}$ and such that $\rho_{\vec a}^{(m)}(c)=0$. The 4D $\CN=2$ SYM theories can also be obtained from a brane construction, this time using type IIA string theory \cite{Witten1997}. In this construction, D5 and NS5 branes are replaced by D4 and solitonic 5-branes respectively. The horizontal representation will be associated to the solitonic 5-branes, while stacks of $m$ D4-branes, on which act a $U(m)$ gauge group, will be associated to the rank $m$ vertical representation. The weights of representations will be identified with the branes positions. In addition, automorphisms of the DIM algebra encode the invariance of the brane-web under geometric transformations \cite{Bourgine2018a}. These automorphisms can also be defined on the algebra \ref{algebra}, except for Miki's automorphism $\CS$ \cite{Miki2007} corresponding to a $90^\circ$ rotation of the brane-web that maps D5 on NS5 and vice versa, thus realizing the S-duality of type IIB string theory. On the other hand, the 180$^\circ$ rotation $\CS^2$ is preserved:
\begin{align}\label{def_autom}
\begin{split}
&\CS^2\cdot c=-c,\quad \CS^2\cdot x^\pm(z)=-x^\mp(-z),\quad \CS^2\cdot\psi^\pm(z)=\psi^\mp(-z),\\
&\t_\o\cdot c=c,\quad \t_\o\cdot x^\pm(z)=x^\pm(z+\o),\quad\t_\o\cdot\psi^\pm(z)=\psi^\pm(z+\o),\\
&\bt_\bo\cdot c=c,\quad \bt_\bo\cdot(x^\pm(z))=\bo^{\pm1}x^\pm(z),\quad \bt_\bo\cdot\psi^\pm(z)=\psi^\pm(z).
\end{split}
\end{align}
Here $\t_\o$ and $\bt_\bo$ define translations in the directions corresponding to the solitonic 5-branes and the D4-branes respectively. In addition, two reflection symmetries $\s_V$ and $\s_H$ (resp. along the directions of D5/D4-branes and NS5/solitonic 5-branes) were introduced in \cite{Bourgine2017b,Bourgine2017c} as involutive isomorphisms of algebra. Their definition extend to the algebra \ref{algebra}, mapping it to the same algebra but with the opposite parameters $\e_\a\to-\e_\a$, and acting on the generators as 
\begin{align}
\begin{split}
&\s_V(c)=c,\quad \s_V\cdot x^\pm(z)=-x^\mp(z),\quad \s_V\cdot \psi^\pm(z)=\psi^\pm(z),\\
&\s_H(c)=-c,\quad \s_H\cdot x^\pm(z)=x^\pm(-z),\quad \s_H\cdot \psi^\pm(z)=\psi^\mp(-z).
\end{split}
\end{align}
Note that $\CS^2$ is also involutive, and $\s_V\s_H=\CS^2$.

The main difference between \ref{algebra} and the DIM algebra is the absence of a mode expansion for the currents $x^\pm(z)$ and $\psi^\pm(z)$. This is due to the presence of zero modes in the horizontal representation, but also to the fact that we use the Dirac delta function instead of the multiplicative one. In practice, the algebra \ref{algebra} is simply a convenient way to combine the algebraic relations satisfied by the currents in the two types of representations. Yet, its main interest lies in the possibility to define a co-algebraic structure and, later on, the corresponding intertwiners. The coproduct can be seen as a degenerate limit of the Drinfeld coproduct for the DIM algebra,
\begin{align}
\begin{split}\label{def_D}
&\D(x^+(z))=x^+(z)\otimes1+\psi^-(z-\ep c_{(1)}/4)\otimes x^+(z-\ep c_{(1)}/2),\\
&\D(x^-(z))=1\otimes x^-(z)+x^-(z-\ep c_{(2)}/2)\otimes\psi^+(z-\ep c_{(2)}/4),\\
&\D(\psi^\pm(z))=\psi^\pm(z\mp\ep c_{(2)}/4)\otimes\psi^\pm(z\pm\ep c_{(1)}/4),
\end{split}
\end{align}
$\D(c)=c_{(1)}+c_{(2)}$ with $c_{(1)}=c\otimes 1$ and $c_{(2)}=1\otimes c$. In fact, the whole Hopf algebra structure survives the limit, the co-unit taking the usual form $\epsilon(x^\pm(z))=\epsilon(c)=0$, $\epsilon(\psi^\pm(z))=\epsilon(1)=1$, and the antipode $\th(c)=-c$, $\th(\psi^\pm(z))=\psi^\pm(z)^{-1}$ and
\begin{equation}
\th(x^+(z))=-\psi^-(z+\e_+c/4)^{-1}x^+(z+\e_+c/2),\quad \th(x^-(z))=-x^-(z+\ep c/2)\psi^+(z+\ep c/4)^{-1}.
\end{equation}
As in \cite{Bourgine2018a}, the opposite (or permuted) coproduct $\D'$ can also be obtained as a twist of $\D$ by the automorphism $\CS^2$.

\subsection{Horizontal representation}
The horizontal representation is built using vertex operators acting on the Fock space of a free boson. The modes $\a_n$ and $P$, $Q$ satisfy the standard Heisenberg commutation relations $[\a_n,\a_m]=n\d_{n+m}$ and $[P,Q]=1$. The vacuum $\ket{\vac}$ is annihilated by the positive modes $\a_{n>0}$ and $P$, while negative modes $\a_{n<0}$ and $Q$ create excitations. As usual the normal ordering $:\cdots:$ is defined by moving the positive modes to the right. We also introduce the dual state $\bra{\vac}$ annihilated by negative modes, so that $\bra{\vac}:\cdots:\ket{\vac}=1$. In our construction, and just like in DIM's horizontal representations \cite{Feigin2009a}, twists between positive and negative modes have to be introduced, and it is convenient to define
\begin{equation}\label{def_vphi}
\vphi_+(z)=P\log z-\sum_{n>0}\dfrac{z^{-n}}{n}\a_n,\quad \vphi_-(z)=Q+\sum_{n>0}\dfrac{z^n}{n}\a_{-n}\implies e^{\vphi_+(z)}e^{\vphi_-(w)}=(z-w):e^{\vphi_+(z)}e^{\vphi_-(w)}:.
\end{equation} 
In fact, the bosonic modes enter in the representation only through the combinations $V_-(z)=e^{\vphi_-(z-\e_2/2)-\vphi_-(z+\e_2/2)}$ and $V_+(z)=e^{\vphi_+(z+\e_1/2)-\vphi_+(z-\e_1/2)}$ that obey the normal-ordering property
\begin{equation}\label{def_S}
V_+(z)V_-(w)=S(z-w-\e_+/2)^{-1}:V_+(z)V_-(w):,\quad\text{with}\quad S(z)=\dfrac{(z+\e_1)(z+\e_2)}{z(z+\e_+)}.
\end{equation} 
The function $S(z)$ is related to the scattering function $g(z)$ appearing in \ref{algebra} by $g(z)=S(z)/S(-z)$, it satisfies the crossing symmetry $S(-\e_+-z)=S(z)$. Then, the horizontal representation of weight $u$ can be defined in terms of the vertex operators
\begin{equation}\label{def_eta}
\eta^\pm(z)=V_-(z\pm\e_+/4)^{\pm1}V_+(z\mp\e_+/4)^{\pm1},\quad \xi^\pm(z)=V_\pm(z\mp\e_+/2)V_\pm(z\pm\e_+/2)^{-1},
\end{equation} 
it reads
\begin{equation}
\rho_u^{(H)}(x^\pm(z))=u^{\pm1}\eta^\pm(z),\quad \rho_u^{(H)}(\psi^\pm(z))=\xi^\pm(z),\quad \rho_u^{(H)}(c)=1.
\end{equation} 
Note that $\xi^+(z)$ (resp. $\xi^-(z)$) is expressed only in terms of positive (negative) modes. The algebraic relations \ref{algebra} follow from the normal-ordering properties of the vertex operators given in \ref{NO_eta_xi}. Here, product of operators are ordered along the imaginary axis: $\CO_1(z)\CO_2(w)$ implies $\Im(z)<\Im(w)$, and the parameters $\e_1$, $\e_2$ are assumed to be real. The only non-trivial relation to be checked is the commutator $[x^+,x^-]$. It follows from the identities $\xi^\pm(z)=:\eta^+(z\mp\e_+/4)\eta^-(z\pm\e_+/4):$, together with pole decomposition of the function
\begin{equation}\label{dec_S}
S(z)=1+\dfrac{\e_1\e_2}{\e_+}\left(\dfrac1z-\dfrac1{z+\e_+}\right)\implies [S(z)]_+-[S(z)]_-=\dfrac{\e_1\e_2}{\e_+}\left(\d(z)-\d(z+\e_+)\right),
\end{equation} 
where $[S(z)]_\pm=S(z\mp i0)$ defines the poles prescription (see \ref{AppA}).

\subsection{Vertical representations and SHc algebra}
Vertical representations of rank $m$ and weight $\vec a=a_1,\cdots,a_m$ act on a basis of states $\dket{\vec a,\vec \l}$ parameterized by a set of $m$ Young diagrams $\vec\l=\l^{(1)},\cdots,\l^{(m)}$,
\begin{align}
\begin{split}\label{def_vert}
&\rho^{(m)}_{\vec a}(x^+(z))\dket{\vec a,\vec \l}=\sum_{x\in A(\vec\l)}\d(z-\phi_x)\res_{z=\phi_x}\CY_{\vec\l}(z)^{-1}\dket{\vec a,\vec \l+x},\\
&\rho^{(m)}_{\vec a}(x^-(z))\dket{\vec a,\vec \l}=\sum_{x\in R(\vec\l)}\d(z-\phi_x)\res_{z=\phi_x}\CY_{\vec\l}(z+\e_+)\dket{\vec a,\vec \l-x},\\
&\rho^{(m)}_{\vec a}(\psi^\pm(z))\dket{\vec a,\vec \l}=\left[\Psi_{\vec\l}(z)\right]_\pm\dket{\vec a,\vec \l}.
\end{split}
\end{align}
We have denoted $A(\vec\l)$ (resp. $R(\vec\l)$) the set of boxes that can be added to (removed from) the Young diagrams composing $\vec\l$. In addition, to each box $x=(l,i,j)\in\vec\l$ with $(i,j)\in\l^{(l)}$, we have associated the complex number $\phi_x=a_l+(i-1)\e_1+(j-1)\e_2$ that is sometimes called \textit{instanton position} (in the moduli space). Finally, the functions $\CY_{\vec\l}(z)$ and $\Psi_{\vec\l}(z)$ appearing in \ref{def_vert} are defined by 
\begin{equation}\label{def_CY}
\Psi_{\vec\l}(z)=\dfrac{\CY_{\vec\l}(z+\e_+)}{\CY_{\vec\l}(z)},\quad \CY_{\vec\l}(z)=\dfrac{\prod_{x\in A(\vec\l)}z-\phi_x}{\prod_{x\in R(\vec\l)}z-\e_+-\phi_x}.
\end{equation} 
Parameters $a_l$ and $\e_1$, $\e_2$ are chosen real, so that the poles lie on the real axis. The commutator $[x^+,x^-]$ in \ref{algebra} follows from the pole decomposition of $\Psi_{\vec\l}(z)$. The other relations are derived using the shell formulas
\begin{equation}
\CY_{\vec\l}(z)=\prod_{l=1}^m(z-a_l)\prod_{x\in\vec\l}S(\phi_x-z),\quad \Psi_{\vec\l}(z)=\prod_{l=1}^m\dfrac{z+\e_+-a_l}{z-a_l}\prod_{x\in\vec\l}g(z-\phi_x).
\end{equation}

In order to define the dual intertwiner, we need to introduce a dual basis $\dbra{\vec a,\vec\l}$ on which acts the contragredient representation $\hat\rho^{(m)}_{\vec a}$ such that
\begin{equation}
\dbra{\vec a,\vec\l}\left(\rho^{(m)}_{\vec a}(e)\dket{\vec a,\vec\mu}\right)=\left(\dbra{\vec a,\vec\l}\hat\rho^{(m)}_{\vec a}(e)\right)\dket{\vec a,\vec\mu}.
\end{equation} 
The form of the contragredient representation depends on the definition of the scalar product between vertical states. Like in the case of DIM's vertical representations \cite{Bourgine2017c,Bourgine2017b}, and in contrast with our previous choice \cite{Bourgine2015c}, it is convenient to introduce a non-trivial norm for these states, namely
\begin{equation}\label{def_al}
\dbra{\vec a,\vec\l}\vec a,\vec\mu\rangle\!\rangle=\Zv(\vec a,\vec\l)^{-1}\d_{\vec\l,\vec\mu},
\end{equation} 
where $\Zv(\vec a,\vec\l)$ denotes the vector contribution to the $\CN=2$ instanton partition function.\footnote{We remind here the formula for the contributions to the instanton partition functions of the multiplets vector $\Zv(\vec a,\vec\l)$ and bifundamental (with mass $\mf$) $\Zbf(\vec a,\vec\l;\vec b,\vec\mu|\mf)$: $\Zv(\vec a,\vec\l)=\Zbf(\vec a,\vec\l;\vec a,\vec\l|0)^{-1}$ and
\begin{equation}
\Zbf(\vec a,\vec\l;\vec b,\vec\mu|\mf)=\prod_{x\in\vec\l}\prod_{l}(\phi_x-b_l+\e_+-\mf)\times\prod_{x\in\vec\mu}\prod_{l}(\phi_x-a_l+\mf)\times\prod_{\superp{x\in\vec\l}{y\in\vec\mu}}S(\phi_x-\phi_y-\mf).
\end{equation}} With this choice, the contragredient representation reads exactly as in \ref{def_vert}, but with the action of $x^\pm(z)$ and $-x^\mp(z)$ exchanged. At first encounter, this choice might look like an artificial way to introduce the vector contribution. However, the underlying algebraic structure is in fact very rigid, and had we chosen a trivial norm, we would have recovered the vector contribution $\Zv(\vec a,\vec\l)$ in the normalization factor $t_{\vec\l}^\ast$ of the dual intertwiner $\Phi^\ast$ defined below. Indirectly here, we have exploited the reflection symmetry $\s_V$ in order to simply the expression of $\Phi^\ast$.

The connection between the vertical representation \ref{def_vert} and the action of SHc algebra on the basis $\dket{\vec a,\vec\l}$ follows from the decomposition $\rho^{(m)}_{\vec a}(x^\pm(z))=[X^\pm(z)]_+-[X^\pm(z)]_-$ and $\psi^\pm(z)=[\Psi(z)]_\pm$ with 
\begin{align}\label{act_Xpm}
\begin{split}
&X^+(z)\dket{\vec a,\vec \l}=\sum_{x\in A(\vec\l)}\dfrac1{z-\phi_x}\res_{z=\phi_x}\CY_{\vec\l}(z)^{-1}\dket{\vec a,\vec \l+x},\\
&X^-(z)\dket{\vec a,\vec \l}=\sum_{x\in R(\vec\l)}\dfrac1{z-\phi_x}\res_{z=\phi_x}\CY_{\vec\l}(z+\e_+)\dket{\vec a,\vec \l-x},
\end{split}
\end{align}
and $\Psi(z)\dket{\vec a,\vec \l}=\Psi_{\vec\l}(z)\dket{\vec a,\vec \l}$. This operator can be expressed using the \textit{half-boson} (i.e. containing only commuting positive modes) $\Phi(z)$,
\begin{equation}\label{def_Phi}
\Psi(z)=\prod_{l=1}^m\dfrac{z+\e_+-a_l}{z-a_l}\prod_{\a=1,2,3}e^{\Phi(z+\e_\a)-\Phi(z-\e_\a)},\quad \Phi(z)=\log(z)\Phi_0-\sum_{n=1}^\infty\dfrac1{nz^n}\Phi_{n},
\end{equation} 
which acts diagonally on the states $\dket{\vec v,\vec\l}$,
\begin{equation}\label{act_Phi}
\Phi_n\dket{\vec v,\vec\l}=\left(\sum_{x\in\vec\l}\phi_x^n\right)\dket{\vec v,\vec\l},\quad \p_z\Phi(z)\dket{\vec v,\vec\l}=\left(\sum_{x\in\vec\l}\dfrac1{z-\phi_x}\right)\dket{\vec v,\vec\l}.
\end{equation} 
The new operators $X^\pm(z)$, $\Psi(z)$ and $\p_z\Phi(z)$ obey the algebraic relations
\begin{equation}
[X^+(z),X^-(w)]=\dfrac{\e_1\e_2}{\e_+}\dfrac{\Psi(z)-\Psi(w)}{z-w},\quad [\p_z\Phi(z),X^\pm(w)]=\pm\dfrac1{z-w}X^\pm(w),
\end{equation} 
leading to identify them with the SHc currents $\sqrt{-\e_1\e_2}D_{\pm1}(z)$, $E(z)$ and $D_0(z)$ (respectively) in the holomorphic presentation of reference \cite{Bourgine2015c}. Furthermore, the actions \ref{act_Xpm} and \ref{act_Phi} on the states $\dket{\vec v,\vec\l}$ coincide with the one presented in \cite{Bourgine2015c}, up to a change of states normalization.\footnote{Alternatively, it is possible to expand the currents at infinity, thus defining the modes
\begin{equation}
X^\pm(z)=\sum_{n=0}^\infty z^{-n-1}X_n^\pm,\quad \Psi(z)=1+\e_+\sum_{n=0}^\infty z^{-n-1}\Psi_n,
\end{equation} 
and identify $X_n^\pm$, $\Psi_n$ and $\Phi_n$ with, respectively, the representation of the SHc generators $\sqrt{-\e_1\e_2}D_{\pm1,n}$, $E_n$ and $D_{0,n+1}$. The latter satisfy the algebraic relations $[D_{0,n},D_{0,m}]=0$, $[D_{0,n+1},D_{\pm1,m}]=\pm D_{\pm1,n+m}$, $[D_{-1,n},D_{1,m}]=E_{n+m}$ and a certain exponential relation between $E_n$ and $D_{0,n}$ deriving from \ref{def_Phi}.}

\section{Reconstructing the gauge theories' BPS quantities}
\subsection{Intertwiners}
In the algebraic engineering of 5D $\CN=1$ gauge theories, the role of the topological vertex is devoted to the intertwiners of the DIM algebra, first introduced by Awata, Feigin and Shiraishi (AFS) in \cite{Awata2011}, and further generalized in \cite{Bourgine2017b,Bourgine2018a}. In the 4D case, the intertwiners, denoted $\Phi^{(m)}[u,\vec a]$ and $\Phi^{\ast(m)}[u,\vec a]$, are obtained as solution to the following equations,
\begin{align}
\begin{split}\label{AFS_lemmas}
&\rho_{u'}^{(H)}(e)\Phi^{(m)}[u,\vec a]=\Phi^{(m)}[u,\vec a]\ \left(\rho_{\vec a}^{(m)}\otimes\rho_{u}^{(H)}\ \D(e)\right),\\
&\left(\rho_{\vec a}^{(m)}\otimes\rho_{u}^{(H)}\ \D'(e)\right)\ \Phi^{\ast(m)}[u,\vec a]=\Phi^{\ast(m)}[u,\vec a]\rho_{u'}^{(H)}(e),
\end{split}
\end{align}
where $e$ is any of the four currents defining the algebra \ref{algebra}, $\D$ is the coproduct given in \ref{def_D}, and $\D'$ the opposite coproduct obtained by permutation. Note that the charge $c$ is conserved in these relations. For each equation, the solution can be expanded on the vertical basis, their components are operators acting on the horizontal Fock space:
\begin{align}
\begin{split}\label{sol_AFS}
&\Phi^{(m)}[u,\vec a]=\sum_{\vec\l}\Zv(\vec a,\vec\l)\ \Phi^{(m)}_{\vec\l}[u,\vec a]\ \dbra{\vec a,\vec\l},\quad \Phi^{(m)}_{\vec\l}[u,\vec a]=t^{(m)}_{\vec\l}[u,\vec a]\ :\prod_{l=1}^m\Phi_{\vac}(a_l)\prod_{x\in\vec\l}\eta^+(\phi_x):,\\
&\Phi^{\ast(m)}[u,\vec a]=\sum_{\vec\l}\Zv(\vec a,\vec\l)\ \Phi^{\ast(m)}_{\vec\l}[u,\vec a]\ \dket{\vec a,\vec\l},\quad \Phi^{\ast(m)}_{\vec\l}[u,\vec a]=t^{\ast(m)}_{\vec\l}[u,\vec a]\ :\prod_{l=1}^m\Phi_{\vac}^{\ast}(a_l)\prod_{x\in\vec\l}\eta^-(\phi_x):,
\end{split}
\end{align}
where $t^{(m)}_{\vec\l}[u,\vec a]=(-1)^{m|\vec\l|}u^{|\vec\l|}$ and $t^{\ast(m)}_{\vec\l}[u,\vec a]=u^{-|\vec\l|}$ are normalization coefficients, and $|\vec\l|$ denotes the total number of boxes contained in $\vec\l$. In addition, the horizontal weights must also obey the constraint $u'=(-1)^mu$. The intertwiners \ref{sol_AFS} taken at $m=1$ define a degenerate version of the refined topological vertex, in the AFS free field presentation \cite{Awata2011}, that can be employed to construct directly 4D $\CN=2$ gauge theories.

The definition of the vacuum components $\Phi_\vac(a)$ and $\Phi_\vac^\ast(a)$ in \ref{sol_AFS} requires some explanation. Physically, these operators define a sort of Fermi sea for which $\eta^\pm$ creates excitations. Thus, and by analogy with AFS intertwiners, we would like to define them as a (normal-ordered) infinite product of the operators $\eta^\pm(\phi_x)^{-1}$ over the boxes of a fully filled Young diagram, i.e. $\phi_x=a+(i-1)\e_1+(j-1)\e_2$ with $i,j=1\cdots\infty$. However, in contrast with the DIM scenario, it is not possible to express the resulting operators as vertex operators built upon the modes $\a_n$. We propose two different approaches to deal with this issue. The first one is to cut-off the size of the infinite Young diagram, effectively regularizing the double product to $i,j=1\cdots N$, and sending $N$ to infinity at the end of the calculation. This approach is a little cumbersome as it introduces extra factors that are eventually canceled. Alternatively, it is possible to introduce an auxiliary free boson $\tphi(z)$ such that $\vphi_-(z)=\tphi_-(z-\e_1/2)-\tphi_-(z+\e_1/2)$ and $\vphi_+(z)=\tphi_+(z+\e_2/2)-\tphi_+(z-\e_2/2)$. Assuming that the propagator of the boson $\tphi(z)$ is given by (minus) the log of a function $F(z)$ (conveniently shifted here by $+\e_+/2$), the relations between the positive/negative modes of the bosons $\vphi$ and $\tphi$ imply a certain functional relation for the propagator:\footnote{Note however, that the asymptotic expansion at $z=\infty$ of the function $\G_2(z)$ (see, for instance, reference \cite{Spreafico}),
\begin{equation}\label{exp_G2}
\log\G_2(z)\sim-\dfrac{z^2}{2\e_1\e_2}\left(\log(z)-\dfrac32\right)+\dfrac{\e_+}{2\e_1\e_2}z\left(\log(z)-1\right)+O(\log(z))
\end{equation} 
is not compatible with the usual mode expansion \ref{def_vphi} of a free boson. To resolve this, it is possible to modify the expression of the propagator, defining $F(z)=\th_2(z)\G_2(z)$ where $\th_2(z)$ obeys the same functional equation as $\G_2(z)$ but with $1$ instead of $z$ in the RHS, and possesses an asymptotic expansion that cancels the unwanted terms in \ref{exp_G2}. However, it turns out that the one-loop determinant contributing to the partition function \cite{Losev2003} is usually regularized using the function $\G_2(z)$ \cite{Bourgine2017}. Another possibility is to add extra zero-modes to the boson $\tphi(z)$ to generate the terms $z^2\log z$, $z\log z$,...}
\begin{equation}
e^{\tphi(z)}e^{\tphi(w)}=F(z-w+\e_+/2)^{-1}:e^{\tphi(z)}e^{\tphi(w)}:,\quad \dfrac{F(z+\e_1)F(z+\e_2)}{F(z)F(z+\e_+)}=z.
\end{equation}
This relation can be solved using the double-gamma function, $F(z)=\G_2(z)$ with pseudo-periods $\e_1$ and $\e_2$ (see \cite{Spreafico,Bourgine2017}). Defining 
\begin{equation}\label{def_Phi_vac}
\Phi_\vac(a)=e^{-\tphi_-(a-\e_+/4)}e^{-\tphi_+(a-3\e_+/4)},\quad \Phi_\vac^{\ast}(a)=e^{\tphi_-(a-3\e_+/4)}e^{\tphi_+(a-\e_+/4)},
\end{equation}
all the necessary normal-ordering relation can now be derived, they have been gathered in appendix \ref{NO_Phi_vac}. Adapting the method described in the appendix C of \cite{Bourgine2017b}, they can be used to show that the expressions given in \ref{sol_AFS} indeed solve the equations \ref{AFS_lemmas}.

\subsection{Instanton partition functions and qq-characters}
From this point, we can repeat the construction presented in \cite{Bourgine2017b} for the instanton partition functions and qq-characters of linear quivers gauge theory. For the purpose of illustration, we present here the main results concerning ($A_1$ quiver) pure $U(m)$ gauge theories. The corresponding $\CT$-operators are obtained by coupling an intertwiner $\Phi^{(m)}[u,\vec a]$ to its dual $\Phi^{\ast(m)}[u^\ast,\vec a]$ through a scalar product $\cdot$ in their common vertical representation,
\begin{equation}\label{def_T_A1}
\CT_{U(m)}=\Phi^{(m)}[u,\vec a]\cdot\Phi^{\ast(m)}[u^\ast,\vec a]=\sum_{\vec\l}\Zv(\vec a,\vec\l)\ \Phi^{(m)}_{\vec\l}[u,\vec a]\otimes \Phi^{\ast(m)}_{\vec\l}[u^\ast,\vec a].
\end{equation}
This operator acts on the tensor product of two bosonic Fock spaces, it is easy to check that its vev reproduces the instanton partition function of the gauge theory,
\begin{equation}
\Zinst[U(m)]=\left(\bra{\vac}\otimes\bra{\vac}\right)\CT_{U(m)}\left(\ket{\vac}\otimes\ket{\vac}\right)=\sum_{\vec\l}\qf^{|\vec\l|}\Zv(\vec a,\vec\l),
\end{equation}
under the identification of $\vec a$ with the Coulomb branch vevs, and $\qf=(-1)^m u/u^\ast$ with the exponentiated gauge coupling. The partition function can also be recovered as a scalar product of Gaiotto states, corresponding here to the vevs of the intertwiners: 
\begin{align}
\begin{split}
&\dket{G,\vec a}=\left(1\otimes\bra{\vac}\right)\Phi^{\ast(m)}[u^\ast,\vec a]\ket{\vac}=\sum_{\vec\l}\Zv(\vec a,\vec\l)\ \bra{\vac}\Phi^{\ast(m)}_{\vec\l}[u^\ast,\vec a]\ket{\vac}\ \dket{\vec a,\vec\l},\\
&\dbra{G,\vec a}=\bra{\vac}\Phi^{(m)}[u,\vec a]\left(1\otimes\ket{\vac}\right)=\sum_{\vec\l}\Zv(\vec a,\vec\l)\ \bra{\vac}\Phi^{(m)}_{\vec\l}[u,\vec a]\ket{\vac}\ \dbra{\vec a,\vec\l},
\end{split}
\end{align}
so that $\la\CT_{U(m)}\ra=\dbra{G,\vec a}\!G,\vec a\rangle\!\rangle$. In fact, these Gaiotto states coincide with those defined in \cite{Bourgine2015c} if we take into account the different normalization of the basis. The action of SHc currents $X^\pm(z)$ on these states has been derived in \cite{Bourgine2015c}. It can be recovered here using the intertwining property \ref{AFS_lemmas}. For instance,
\begin{align}
\begin{split}
&\rho_{\vec a}^{(m)}(x^+(z))\dket{G,\vec a}=-u^\ast\left([\CY(z+\e_+)_+-[\CY(z+\e_+)]_-\right)\dket{G,\vec a},\\
&\rho_{\vec a}^{(m)}(x^-(z))\dket{G,\vec a}=-(u^\ast)^{-1}\left([\CY(z)^{-1}]_+-[\CY(z)^{-1}]_-\right)\dket{G,\vec a},
\end{split}
\end{align}
where $\CY(z)$ is a diagonal operator in the basis $\dket{\vec a,\vec\l}$ with eigenvalues $\CY_{\vec\l}(z)$.

By construction, the operator $\CT_{U(m)}$ commutes with the action of any element $e$ of the algebra \ref{algebra} in the appropriate representation,
\begin{equation}\label{covar_T}
\left(\rho_{u'}^{(H)}\otimes\rho_{u^\ast}^{(H)}\ \D'(e)\right)\CT_{U(m)}=\CT_{U(m)}\left(\rho_{u}^{(H)}\otimes\rho_{u^{\ast\prime}}^{(H)}\ \D'(e)\right).
\end{equation}
The fundamental qq-character can be obtained by insertion of the current $x^+(z)$ before taking the vev,\footnote{Note that, as consequence of the invariance of the brane web under the reflection encoded by $\s_V$, we find the same quantity if we replace $x^+(z)$ by $x^-(z)$ (and $(u^\ast)^{-1}$ by $u'$) in this formula.}
\begin{equation}\label{def_chi}
\chi(z-\e_+/2)=(u^\ast)^{-1}\dfrac{\left(\bra{\vac}\otimes\bra{\vac}\right)\left(\rho_{u'}^{(H)}\otimes\rho_{u^\ast}^{(H)}\ \D'(x^+(z))\right)\CT_{U(m)}\left(\ket{\vac}\otimes\ket{\vac}\right)}{\left(\bra{\vac}\otimes\bra{\vac}\right)\CT_{U(m)}\left(\ket{\vac}\otimes\ket{\vac}\right)}.
\end{equation}
After evaluation of the vev using the normal-ordering relations \ref{NO_Intw} of \ref{AppB}, we recover the well-known expression
\begin{equation}\label{expr_chi}
\chi(z)=\Zinst[U(m)]^{-1}\sum_{\vec\l}\qf^{|\vec\l|}\Zv(\vec a,\vec\l)\left(\CY_{\vec\l}(z+\e_+)+\dfrac{\qf}{\CY_{\vec\l}(z)}\right).
\end{equation}
The covariance property \ref{covar_T} of the operator $\CT_{U(m)}$ implies the cancellation of poles between the two terms in \ref{expr_chi}, and so $\chi(z)$ is a polynomial of degree $m$. Higher qq-characters are constructed by insertion of multiple operators $x^+(z_i)$ in the vevs.

These results can be extended to linear quivers of higher rank. The corresponding $\CT$-operator is obtained by coupling several of the operators $\CT_{U(m)}$ defined in \ref{def_T_A1} in the horizontal channel. Bifundamental contributions $\Zbf$ to the partition function appear as the product of intertwiners are normal-ordered following the two relations
\begin{align}\small
\begin{split}
&\Phi^{(m)}_{\vec\l}[u,\vec a]\Phi^{\ast(m^\ast)}_{\vec\l^\ast}[u^\ast,\vec a^\ast]=(-1)^{m|\vec\l^\ast|}\Zbf(\vec a,\vec\l;\vec a^\ast,\vec\l^\ast|\e_+/2)\prod_{l,l^\ast=1}^{m,m^\ast}F(a_l-a_{l^\ast}^\ast+\e_+/2)\ :\Phi^{(m)}_{\vec\l}[u,\vec a]\Phi^{\ast(m^\ast)}_{\vec\l^\ast}[u^\ast,\vec a^\ast]:,\\
&\Phi^{\ast(m^\ast)}_{\vec\l^\ast}[u^\ast,\vec a^\ast]\Phi^{(m)}_{\vec\l}[u,\vec a]=(-1)^{m^\ast|\vec\l|}\Zbf(\vec a,\vec\l;\vec a^\ast,\vec\l^\ast|\e_+/2)\prod_{l,l^\ast=1}^{m,m^\ast}F(a_{l^\ast}^\ast-a_l+\e_+/2)\ :\Phi^{(m)}_{\vec\l}[u,\vec a]\Phi^{\ast(m^\ast)}_{\vec\l^\ast}[u^\ast,\vec a^\ast]:.
\end{split}
\end{align}
In these formulas, the $F$-dependent factor can be interpreted as one-loop bifundamental contributions with the $\G_2$-regularization of infinite products. The fundamental qq-characters attached to the two nodes at the extremity of the quiver are obtained by insertion of $x^\pm(z)$ (with repeated coproduct). On the other hand, qq-characters corresponding to middle nodes require the introduction of more involved operators \cite{Bourgine2017b}.

\subsection{Degenerate limit of Kimura-Pestun's quiver W-algebra}
Instanton partition functions of 5D $\CN=1$ quiver gauge theories are covariant under the action of two distinct q-deformed W-algebras. The first one is involved in the q-deformed AGT correspondence \cite{Awata2009}, it appears at each individual node of the quiver. For instance, the partition function of a linear quiver theory bearing a $U(m)$ gauge group at each node coincides with a conformal block of the q-W$_m$-algebra. A second form of q-W-algebra covariance has appeared in the work of Kimura and Pestun (KP) \cite{Kimura2015}. In their construction, the Dynkin diagram of the algebra is identified with the quiver of the gauge theory. Thus, the partition function of a linear quiver with $k$ nodes is covariant under the action of a q-W$_{k+1}$-algebra. The DIM algebra offers a third perspective which is somehow more elementary as it allows us to understand the relation between these two q-W-algebras (at least in the case of linear quivers). For instance, each node of a quiver bearing a $U(m)$ gauge group corresponds to a stack of $m$ D5-branes in the $(p,q)$-brane construction. Accordingly, the node carries a vertical representation $(0,m)$ that contains the q-$W_m$ action involved in the q-AGT correspondence. On the other hand, the DIM algebra acts on the $\CT$-operators in a different representation, obtained as a tensor product of horizontal representations $(1,n)$ attached to the external NS5-branes (possibly dressed by D5s). This tensored horizontal representation factorizes into a q-Heisenberg contribution, and the action of KP's q-W-algebra \cite{Mironov2016,Bourgine2017b}. Furthermore, the automorphism $\CS$ discovered by Miki \cite{Miki2007} maps vertical to horizontal representations (and vice-versa), effectively implementing a rotation of the $(p,q)$-brane web, and thereby exchanging the role of the two different q-W-algebras. We refer to the upcoming paper \cite{Bourgine2018a} for a deeper discussion of this automorphism in the formalism of algebraic engineering.

In this section, we would like to define the degenerate version of KP's quiver W-algebra that acts on partition functions of 4D $\CN=2$ gauge theories. We will restrict ourselves to the case of a single node quiver with $U(m)$ gauge group. For the 5D version of the gauge theory, the relevant q-W-algebra is the q-Virasoro algebra, constructed in \cite{Shiraishi1995} as a q-deformation of the usual Virasoro algebra. In the DIM description, the representation acting on the $\CT$-operator has levels $(2,0)$, it is obtained by coproduct of the two representations $(1,n)\otimes(1,-n)$ associated to the two (dressed) NS5-branes. It can be decomposed as a tensor product of KP's q-Virasoro algebra and a q-Heisenberg algebra \cite{Mironov2016,Bourgine2017b}. The latter encodes the action of the currents $\bar\psi^\pm(Z)$, it can be factorized out of the action of $\bar x^\pm(Z)$, thereby providing the action of q-Virasoro on the $\CT$-operator. Here, we would like to repeat this construction for the degenerate DIM algebra \ref{algebra}.

The first step is to isolate the q-Heisenberg action. For this purpose we need to introduce an auxiliary bosonic field $\vphi_W(z)$, with propagator $\log h(z)$, such that the (half) vertex operators $V_\pm(z)$ are decomposed into products of $W_\pm(z)=e^{\vphi_{W\pm}(z)}$:\footnote{In terms of bosonic fields, we have the relations $\vphi_{W+}(z+\e_+/2)+\vphi_{W+}(z-\e_+/2)=\vphi_+(z+\e_1/2)-\vphi_+(z-\e_1/2)$ and $\vphi_{W-}(z+\e_+/2)+\vphi_{W-}(z-\e_+/2)=\vphi_-(z-\e_2/2)-\vphi_-(z+\e_2/2)$ for positive/negative modes.}
\begin{equation}\label{dec_V_W}
V_\pm(z)=W_\pm(z+\e_+/2)W_\pm(z-\e_+/2),\quad W_+(z)W_-(w)=h(z-w):W_+(z)W_-(w):.
\end{equation}
It turns out convenient to introduce also the function $f(z)=h(z-\e_+/2)^{-1}h(z+\e_+/2)^{-1}$. Since the decomposition \ref{dec_V_W} has to reproduce the normal-ordering property \ref{def_S} for the vertex operators $V_\pm(z)$, we find the functional relation $f(z)f(z+\e_+)=S(z)$ constraining $f(z)$ in terms of the function $S(z)$. The exact formula for the function $f(z)$ will be specified later. Using the decomposition \ref{dec_V_W}, it is easy to observe that the representation $\rho^{(H)}\otimes\rho^{(H)}$ of the currents $x^\pm(z)$, $\psi^\pm(z)$ can be written as\footnote{It is possible to work with either coproduct $\D$ or $\D'$ by simply permuting the two Fock spaces. Here we chose $\D'$ in agreement with the previous section.}
\begin{align}\label{dec_UT}
\begin{split}
&\rho_{u_1}^{(H)}\otimes\rho_{u_2}^{(H)}\ \D'(x^\pm(z))=\sqrt{u_1u_2}^{\pm1}U_-(z\pm\e_+/2)^{\pm1}T(z)U_+(z\mp\e_+/2)^{\pm1},\\ &\rho_{u_1}^{(H)}\otimes\rho_{u_2}^{(H)}\ \D'(\psi^\pm(z))=U_\pm(z\mp\e_+)U_\pm(z\pm\e_+)^{-1},
\end{split}
\end{align}
where we have introduced the following quantities:
\begin{align}
\begin{split}
&U_\pm(z)=W_\pm(z\pm\e_+/4)\otimes W_\pm(z\mp\e_+/4),\quad T(z)=R(z)+:R(z-\e_+)^{-1}:,\\
&R(z)=\sqrt{\dfrac{u_2}{u_1}}W_-(z+\e_+/4)^{-1}W_+(z-\e_+/4)^{-1}\otimes W_-(z-\e_+/4)W_+(z+\e_+/4).
\end{split}
\end{align}

It is possible to refine the decomposition \ref{dec_UT}, observing that the vertex operators $U_\pm(z)$ and $R(w)$ commute as a result of the cancellation between the factors obtained in each channel when normal-ordering. Commuting the factors of $U_\pm(z)$ with $T(z)$, we can write the representation of the currents $x^\pm(z)$ and $\psi^\pm(z)$ in the form
\begin{equation}
\rho_{u_1}^{(H)}\otimes\rho_{u_2}^{(H)}\ \D'(x^\pm(z))=V_B^\pm(z)T(z),\quad \rho_{u_1}^{(H)}\otimes\rho_{u_2}^{(H)}\ \D'(\psi^\pm(z))=:V_B^+(z\mp\e_+/2)V_B^-(z\pm\e_+/2):,
\end{equation}
where $V_B^\pm(z)=\sqrt{u_1u_2}^{\pm1}U_-(z\pm\e_+/2)^{\pm1}U_+(z\mp\e_+/2)^{\pm1}$ commutes with $T(w)$. As in the case of the DIM algebra, we observe that the action of the currents $\psi^\pm(z)$ is given only in terms of the vertex operators $V_B^\pm(z)$. These operators can be factorized out of the action of $x^\pm(z)$, leaving only the operator $T(z)$. The latter obeys the following weighted
commutation relation,\footnote{This relation is obtained by first computing the normal-ordering
\begin{align}
\begin{split}
T(z)T(w)&=f(z-w)^{-1}\left[:R(z)R(w):+:R(z-\e_+)^{-1}R(w-\e_+)^{-1}:\right]+f(z-w+\e_+):R(z)R(w-\e_+)^{-1}:\\
&+f(z-w-\e_+):R(z-\e_+)^{-1}R(w):,
\end{split}
\end{align}
using \ref{dec_V_W}. Multiplying this relation with $f(z-w)$, the function $S(z-w)$ appears in the RHS. In order to recover \ref{rel_T}, we take the difference between the two normal-ordering relations with $z$ and $w$ exchanged. The first terms cancel, while the remaining ones produce the $\d$-function through the pole decomposition \ref{dec_S} of the function $S(z)$.}
\begin{equation}\label{rel_T}
f(z-w)T(z)T(w)-f(w-z)T(w)T(z)=-\dfrac{\e_1\e_2}{\e_+}\left(\d(z-w+\e_+)-\d(z-w-\e_+)\right),
\end{equation} 
that will be identified with a certain degenerate limit of the q-Virasoro algebra. We claim that this algebra is the degenerate version of KP's quiver W-algebra in the case of a single node. Repeating this procedure for an arbitrary quiver, it should be possible to formulate the general degenerate KP algebras.

The q-Virasoro algebra is defined in \cite{Shiraishi1995} in terms of its \textit{stress-energy tensor} $\bT(Z)$, a current that obeys the algebraic relation
\begin{equation}\label{rel_bT}
\bf(W/Z)\bT(Z)\bT(W)-\bf(Z/W)\bT(W)\bT(Z)=\dfrac{(1-q_1)(1-q_2)}{(1-q_1q_2)}\left(\bd(Z/q_3W)-\bd(q_3Z/W)\right),
\end{equation} 
where $q_1=q$ and $q_2=t^{-1}$ are two complex parameters, $q_3=q_1^{-1}q_2^{-1}$, $\bd(z)$ is the multiplicative delta function \ref{def_bd}, and the function $\bf(Z)$ writes (the second equality is valid when $|q_3|<1$):
\begin{equation}
\bf(Z)=\exp\left(\sum_{n=1}^\infty\dfrac1n\dfrac{(1-q_1^n)(1-q_2^n)}{(1+q_3^{-n})}Z^n\right)=\dfrac1{1-Z}\prod_{n=0}^\infty\dfrac{(1-q_1^{-1}q_3^{2n}Z)(1-q_2^{-1}q_3^{2n}Z)}{(1-q_1^{-1}q_3^{2n+1}Z)(1-q_2^{-1}q_3^{2n+1}Z)}.
\end{equation}
This function $\bf(Z)$ satisfies the functional relation
\begin{equation}\label{fr_bf}
\bf(Z)\bf(q_3^{-1}Z)=\bar S(Z),\quad \text{with}\quad \bar S(Z)=\dfrac{(1-q_1Z)(1-q_2Z)}{(1-Z)(1-q_1q_2Z)}.
\end{equation} 
In order to recover the relation \ref{rel_T} for \ref{rel_bT}, we need to take a degenerate limit. Just like in the case of DIM (see  \ref{AppA}), we set $q_\a=e^{R\e_\a}$, $Z=e^{Rz}$, $W=e^{Rw}$, and send $R\to0$. Then, $\bar S(Z)$ tends to $S(z)$ and the functional relation \ref{fr_bf} reproduces the one satisfied by the function $f(z)$ (which is in fact also satisfied by $f(-z)$ due to the crossing symmetry $S(-\e_+-z)=S(z)$). Since this functional equation is the only constraint for the (entire) function $f(z)$, we choose to identify $f(z)$ with the limit of $\bf(Z^{-1})$:
\begin{equation}
f(z)=\lim_{R\to0}\bf(e^{-Rz})=\dfrac{z-\e_2}{z}\prod_{n=0}^\infty\dfrac{(z+\e_2+2n\e_+)(z-\e_2+(2n+1)\e_+)}{(z-\e_2+2n\e_+)(z+\e_2+(2n+1)\e_+)}.
\end{equation}
Applying the limiting procedure of \ref{AppA} to treat the delta functions in the RHS of \ref{rel_bT}, it is easy to show that the limit $T(z)$ of $\bT(Z)$ obeys \ref{rel_T}. Note that this degenerate limit is different from the usual one that sends q-Virasoro to Virasoro, and Macdonald to Jack polynomials.\footnote{Presumably, in our limit the variables of the polynomial $X_i=e^{R x_i}$ are also sent to one, so that the Macdonald operator does not reduce to the Calogero-Sutherland Hamiltonian, but instead produces a difference operator
\begin{equation}
D_{q,t}=\sum_i\prod_{j\neq i}\dfrac{q_2^{-1}X_i-X_j}{X_i-X_j}q_1^{X_i\p_{X_i}}\to \sum_i\prod_{j\neq i}\dfrac{x_i-x_j-\e_2}{x_i-x_j}e^{\e_1\p_{x_i}}.
\end{equation}}
This particular limit of the q-Virasoro algebra has already been encountered in the study of dualities for 3D $T[U(N)]$ theories, it provides an additive analogue of the q-Virasoro algebra called \textit{d-Virasoro} \cite{Nedelin2017}. Screening currents, vertex operators and Dotsenko-Fateev representations of conformal blocks were also defined in this paper. Furthermore, a correspondence with conformal blocks of ordinary Virasoro algebra was also observed, it suggests the presence of an analogue of Miki's automorphism in the degenerate case. Indeed, in the self-dual pure $U(2)$ gauge theory, the exchange of vertical/horizontal representations induced by the automorphism should reduce to an exchange of the role played by the degenerate KP algebra (d-Virasoro) and the ordinary Virasoro. We hope to come back to this problem in a near future.

Interestingly, the d-Virasoro algebra is also related to the 2D (massive) sine-Gordon model describing the scaling limit of the XYZ model.\footnote{This connection is a priori unrelated with the one observed in \cite{Mironov2009} between pure $SU(2)$ gauge theory and the 1D sine-Gordon model.} Indeed, in the short note \cite{Lukyanov1995}, Lukyanov noticed that the ratio\footnote{We use here the infinite product formula GR.1.431.1 and GR.1431.3 of \cite{Gradshteyn1996} for the trigonometric functions,
\begin{equation}
\sin z=z\prod_{k=1}^\infty\left(1-\dfrac{z^2}{k^2\pi^2}\right),\quad \cos z=\prod_{k=0}^\infty\left(1-\dfrac{4z^2}{(2k+1)^2\pi^2}\right).
\end{equation}}
\begin{equation}
\dfrac{f(z)}{f(-z)}=\dfrac{\tan\left(\pi(z+\e_2)/2\e_+\right)}{\tan\left(\pi(z-\e_2)/2\e_+\right)}
\end{equation}
reproduces the scattering factor of the sine-Gordon model
\begin{equation}
\dfrac{\sinh\b+i\sin\pi\xi}{\sinh\b-i\sin\pi\xi}
\end{equation} 
setting $\b=i\pi z/\e_+$, and $\xi=-\e_2/\e_+$ (note that the ordering $\Im z_1<\Im z_2$ becomes $\Re\b_1>\Re\b_2$). This observation, together with the existence of simple poles at $\b=\pm i\pi$ (corresponding to the delta functions in \ref{rel_T}), led him to identify $T(z)$ with the generator of the corresponding Zamolodchikov-Faddeev (ZF) algebra \cite{Zamolodchikov1978,Faddeev1980}. In this context, the operator $T(\b)$ creates a breather with rapidity $\b$, the basic particle of the theory.\footnote{See also \cite{Fioravanti2005,*Fioravanti2006} for the derivation of the q-Virasoro algebra as the ZF algebra of the first breather in the scaling limit of the spin 1/2 XYZ model using Bethe Ansatz techniques.}


\section{Perspectives}
We have presented here an algebraic construction for the BPS quantities of 4D $\CN=2$ SYM theories based on a degenerate version of the DIM algebra. The two essential ingredients are the action of the SHc algebra (vertical representation), and an algebra of twisted vertex operators (horizontal representation). It would be interesting to understand how the latter fits into the formalism of Maulik-Okounkov \cite{Maulik2012}, and to relate the intertwiners to Smirnov's instanton R-matrix \cite{Smirnov2013}. For linear quivers, the $\CT$-operator constructed here is expected to be the Baxter T-operator of a 1D quantum integrable system, with two spectral parameters associated to vertical and horizontal weights, that produces the set of commuting Hamiltonians by expansion. Given the rich structure of this algebra, there is little doubt that the underlying statistical model will reveal itself of great interest.

Following the program initiated in \cite{Bourgine2014}, it should be possible to take the Nekrasov-Shatashvili limit $\e_2\to0$ of the algebra \ref{algebra}. In this way, one hopes to recover the quantum algebra behind the integrable system of the Bethe/gauge correspondence \cite{Nekrasov2009}. This construction should provide us with a proper interpretation for the $\e_2$-deformation of quantum integrable systems, and reproduce the subleading order terms in $\e_2$ derived in \cite{Bourgine2015,Bourgine2015b} by expansion of the partition function.

Finally, the second main result of this note is the identification of the degenerate version of KP's quiver W-algebra. In the $A_1$ case, we observed an unsuspected connection with the ZF algebra of the sine-Gordon model. This is particularly intriguing given the similarity between the algebraic engineering and the method developed by Lukyanov to compute form factors of massive integrable models \cite{Lukyanov1993}. Indeed, the vertical representation is reminiscent of the ZF algebra's representation $\pi_A$ associated to asymptotic states, while both horizontal representation and Lukyanov's $\pi_Z$ are built upon free fields. It appears essential to understand better the connection between the two methods. In this perspective, one could try to extend our algebraic construction to more general scattering functions $g(z)$. We hope to come back to this issue in a future publication.

\section*{Acknowledgments} 
JEB would like to thank Davide Fioravanti and Vincent Pasquier for discussions, and the IPhT CEA-Saclay, DESY, and the University of Bologna for hospitality at various stages of this project. KZ(Hong Zhang) thanks Wei Li for guidance, and Shuichi Yokoyama for discussions. The work of KZ is partially supported by the General Financial Grant from the China Postdoctoral Science Foundation, with Grant No.
2017M611009.

\appendix
\section{Degenerate limit of DIM algebra}\label{AppA}
In the second Drinfeld presentation, the DIM algebra is defined in terms of four currents $\bar x^\pm(Z)$ and $\bar\psi^\pm(Z)$, and a central element $c$, that obey the algebraic relations,
\begin{align}
\begin{split}\label{def_DIM}
&[\bar\psi^\pm(Z),\bar\psi^\pm(W)]=0,\quad \bar\psi^+(Z)\bar\psi^-(W)=\dfrac{\bar g(q_3^{c/2} Z/W)}{\bar g(q_3^{-c/2}Z/W)}\bar\psi^-(W)\bar\psi^+(Z),\quad \bar x^\pm(Z)\bar x^\pm(W)=\bar g(Z/W)^{\pm1}\bar x^\pm(W)\bar x^\pm(Z),\\
&\bar\psi^+(Z)\bar x^\pm(W)=\bar g(q_3^{\pm c/4}Z/W)^{\pm1}\bar x^\pm(W)\bar\psi^+(Z),\quad \bar\psi^-(Z)\bar x^\pm(W)=\bar g(q_3^{\mp c/4}Z/W)^{\pm1}\bar x^\pm(W)\bar\psi^-(Z),\\
&[\bar x^+(Z),\bar x^-(W)]=\k\left(\bd(q_3^{-c/2}Z/W)\bar\psi^+(q_3^{-c/4}Z)-\bd(q_3^{c/2} Z/W)\bar\psi^-(q_3^{c/4}Z)\right).
\end{split}
\end{align}
The algebra depends on the complex parameters $q_1,q_2,q_3$, with the constraint $q_1q_2q_3=1$, entering in the definition of the $\mathbb{C}$-number $\k$ and the function $\bar g(z)$:
\begin{equation}
\k=\dfrac{(1-q_1)(1-q_2)}{(1-q_1q_2)},\quad \bar g(Z)=\prod_{\a=1,2,3}\dfrac{1-q_\a Z}{1-q_\a^{-1}Z}.
\end{equation}
In order to define the degenerate limit of the algebra, we introduce a parameter $R$, interpreted as the radius of the compact dimension for the 5D background, and write $q_\a=e^{R\e_\a}$, $Z=e^{Rz}$ and $W=e^{Rw}$. The algebra is invariant under the rescaling $\bar x^\pm(Z)\to \bo^{\pm1}\bar x^\pm(Z)$ (the equivalent of the automorphism $\bt_\bo$ in \ref{def_autom}), which leads to some freedom for the definition of the currents' limit:
\begin{equation}
x^\pm(z)=\lim_{R\to0}R^{\pm\a}\bar x^\pm(e^{Rz}),\quad \psi^\pm(z)=\lim_{R\to0}\bar\psi^\pm(e^{Rz}).
\end{equation} 
As $R$ is sent to zero, the function $\bar g(e^{Rz})$ tends to $g(z)$, and the relations \ref{def_DIM} reduce to \ref{algebra}. However, the delta functions entering in the last relation have to be treated carefully. In the DIM algebra, these delta functions are multiplicative ones, they arise from the analytical continuation around a pole at $Z=1$ in the complex plane
\begin{equation}\label{def_bd}
\bd(Z)=\sum_{k\in\mathbb{Z}}Z^k=\left[\dfrac{1}{Z-1}\right]_{\bar +}-\left[\dfrac{1}{Z-1}\right]_{\bar -},
\end{equation} 
where $[f(Z)]_{\bar +}$ denotes the Laurent series in $Z$ obtained after expansion of the function $f(Z)$ for $|Z|>1$ and, $[f(Z)]_{\bar -}$ the series obtained by expansion for $|Z|<1$. Taking the limit $R\to0$, we find that
\begin{equation}
\lim_{R\to0}\left[\dfrac{R}{e^{Rz}-1}\right]_{\bar \pm}=\left[\dfrac{1}{z}\right]_{\pm},
\end{equation} 
where the subscript $\pm$ now has a different meaning which we explain now. For definiteness, we assume that $R$ is a purely positive imaginary number, i.e. $R\in i\mathbb{R}^{>0}$. This corresponds to the choice of an axis for ordering the variables $z$, and we take here the imaginary axis. Then, the condition $|Z|>1$ corresponds to $\Im z<0$ (and obviously $|Z|<1$ to $\Im z>0$). Thus, under the inverse conformal map $z(Z)$, the complex plane is sent to an infinite cylinder of radius $i/R$, and the radial ordering is mapped to a ``\textit{chronological}'' ordering along the imaginary axis identified with the axis of the cylinder. In the limit $R\to0$, the pole at $Z=1$ becomes a pole at $z=0$, and the prescription $[\cdots]_\pm$ should be defined along the imaginary axis,
\begin{equation}
\left[\dfrac{1}{z}\right]_{\pm}=\dfrac1{z\mp i0}\implies \lim_{R\to0}\k\bd(e^{Rz})=-\dfrac{\e_1\e_2}{\ep}\left(\dfrac1{z-i0}-\dfrac1{z+i0}\right)=-\dfrac{\e_1\e_2}{\ep}\d(z).
\end{equation} 

\section{Formulas for operators' normal-ordering}\label{AppB}
In this section, we gather several useful identities for the computation of normal-orderings. In order to write shorter formulas, we will abbreviate the expressions of the type $\CO_1(z)\CO_1(w)=S_{12}(z-w):\CO_1(z)\CO_2(w):$ into $\CO_1(z)\CO_1(w)::S_{12}(z-w)$. Starting with the vertex operators $\eta^\pm(z)$ and $\xi^\pm(z)$ defined in \ref{def_eta}, we find the following non-trivial normal-ordering relations:
\begin{align}\label{NO_eta_xi}
\begin{split}
&\eta^\pm(z)\eta^\pm(w)::S(\mp z\pm w)^{-1},\quad \eta^\pm(z)\eta^\mp(w)::S(z-w-\e_+/2),\quad \xi^+(z)\eta^\pm(w)::g(z-w\mp\e_+/4)^{\pm1},\\
&\eta^\pm(z)\xi^-(w)::g(z-w\mp\e_+/4)^{\pm1},\quad \xi^+(z)\xi^-(w)::\dfrac{g(z-w-\e_+/2)}{g(z-w+\e_+/2)}.
\end{split}
\end{align}
Next, using the definition \ref{def_Phi_vac} of the intertwiners' vacuum components, we find
\begin{align}\label{NO_Phi_vac}\small
\begin{split}
&\eta^+(z)\Phi_\emptyset(a)::(z-a)^{-1},\quad \Phi_\emptyset(a)\eta^+(z)::(a-z-\e_+)^{-1},\quad\eta^-(z)\Phi_\emptyset(a)::(z-a+\e_+/2),\quad \Phi_\emptyset(a)\eta^-(z)::(a-z-\e_+/2),\\
&\xi^+(z)\Phi_\emptyset(a)::\dfrac{z-a+3\e_+/4}{z-a-\e_+/4},\quad \Phi_\emptyset(a)\xi^+(z)::1,\quad \xi^-(z)\Phi_\emptyset(a)::1,\quad \Phi_\emptyset(a)\xi^-(z)::\dfrac{z-a+\e_+/4}{z-a+5\e_+/4},\\
&\eta^+(z)\Phi_\emptyset^\ast(a)::(z-a+\e_+/2),\quad \Phi_\emptyset^\ast(a)\eta^+(z)::(a-z-\e_+/2),\quad\eta^-(z)\Phi_\emptyset^\ast(a)::(z-a+\e_+)^{-1},\quad \Phi_\emptyset^\ast(a)\eta^-(z)::(a-z)^{-1},\\
&\xi^+(z)\Phi_\emptyset^\ast(a)::\dfrac{z-a+\e_+/4}{z-a+5\e_+/4},\quad \Phi_\emptyset^\ast(a)\xi^+(z)::1,\quad \xi^-(z)\Phi_\emptyset^\ast(a)::1,\quad \Phi_\emptyset^\ast(a)\xi^-(z)::\dfrac{z-a+3\e_+/4}{z-a-\e_+/4}.
\end{split}
\end{align}
Finally, we present the relations involving intertwiners:
\begin{align}\label{NO_Intw}\small
\begin{split}
&\eta^+(z)\Phi_{\vec\l}^{(m)}[u,\vec a]::\CY_{\vec\l}(z)^{-1},\quad \eta^-(z)\Phi_{\vec\l}^{(m)}[u,\vec a]::\CY_{\vec\l}(z+\e_+/2),\quad \xi^+(z)\Phi_{\vec\l}^{(m)}[u,\vec a]::\Psi_{\vec\l}(z-\e_+/4),\\
&\eta^+(z)\Phi_{\vec\l}^{(m)\ast}[u,\vec a]::\CY_{\vec\l}(z+\e_+/2),\quad \eta^-(z)\Phi_{\vec\l}^{(m)\ast}[u,\vec a]::\CY_{\vec\l}(z+\e_+)^{-1},\quad \xi^+(z)\Phi_{\vec\l}^{(m)\ast}[u,\vec a]::\Psi_{\vec\l}(z+\e_+/4)^{-1},\\
&\Phi_{\vec\l}^{(m)}[u,\vec a]\eta^+(z)::(-1)^{m}\CY_{\vec\l}(z+\e_+)^{-1},\quad \Phi_{\vec\l}^{(m)}[u,\vec a]\eta^-(z)::(-1)^{m}\CY_{\vec\l}(z+\e_+/2),\quad \Phi_{\vec\l}^{(m)}[u,\vec a]\xi^-(z)::\Psi_{\vec\l}(z+\e_+/4)^{-1}\\
&\Phi_{\vec\l}^{(m)\ast}[u,\vec a]\eta^+(z)::(-1)^{m} \CY_{\vec\l}(z+\e_+/2),\quad \Phi_{\vec\l}^{(m)\ast}[u,\vec a]\eta^-(z)::(-1)^{m} \CY_{\vec\l}(z)^{-1},\quad \Phi_{\vec\l}^{(m)\ast}[u,\vec a]\xi^-(z)::\Psi_{\vec\l}(z-\e_+/4).
\end{split}
\end{align}


\bibliographystyle{utphys_mciteplus}

\end{document}